\documentclass[journal,twoside,shortpaper,9pt]{IEEEtran}
\usepackage{atbegshi,picture,ifpdf,setspace}
\usepackage[caption=false,font=footnotesize]{subfig}
\usepackage[noadjust]{cite}
\usepackage[utf8x]{inputenc}
\usepackage[T1]{fontenc}
\usepackage{url}
\usepackage[detect-all]{siunitx}
\usepackage[english]{babel}
\usepackage[cmex10]{amsmath}
\usepackage{bm}
\usepackage[cmintegrals,varvw]{newtxmath}
\usepackage[pdftex]{graphicx}
\usepackage{booktabs}

\makeatletter
\def\ps@IEEEtitlepagestyle{%
  \def\@oddfoot{\mycopyrightnotice}%
  \def\@oddhead{\hbox{}\@IEEEheaderstyle\leftmark\hfil\thepage}\relax
  \def\@evenhead{\@IEEEheaderstyle\thepage\hfil\leftmark\hbox{}}\relax
  \def\@evenfoot{}%
}
\def\mycopyrightnotice{%
  \begin{minipage}{\textwidth}
  \centering \scriptsize
  Copyright~\copyright~2017 IEEE. Personal use of this material is permitted. Permission from IEEE must be obtained for all other uses, in any current or future media, including\\reprinting/republishing this material for advertising or promotional purposes, creating new collective works, for resale or redistribution to servers or lists, or reuse of any copyrighted component of this work in other works by sending a request to pubs-permissions@ieee.org.
  \end{minipage}
}
\makeatother

\AtBeginShipout{\AtBeginShipoutUpperLeft{%
  \put(\dimexpr\paperwidth-1cm\relax,-.55cm){\makebox[0pt][r]{
  \begin{minipage}{\textwidth}
  \centering \scriptsize 
  This is the author's version of an article that has been published in this journal. 
  Changes were made to this version by the publisher prior to publication. 
  \\[0.3ex]
  The final version of record is available at\qquad http://dx.doi.org/10.1109/TAP.2017.2710261
  \end{minipage}
  }}}%
}

\begin{document}
\title{{\fontsize{24}{26}\selectfont{Communication\rule{29.9pc}{0.675pt}}}\vspace*{0.2cm}\break\fontsize{16}{18}\selectfont
\vspace*{-0.1cm}%
A mm-Wave Patch Antenna with Broad Bandwidth and a Wide Angular Range}
  \author{\fontsize{12}{14}\selectfont Jonas Kornprobst, Kun Wang, Gerhard Hamberger and Thomas F. Eibert%
  \thanks{Manuscript received August 26, 2016; revised April 21, 2017; accepted May 21, 2017. \emph{(Corresponding author: Jonas Kornprobst.)}}
  \thanks{J. Kornprobst, G. Hamberger and T. F. Eibert are with the Chair of High-Frequency Engineering, Department of Electrical and Computer Engineering, Technical University of Munich, Germany (e-mail: hft@ei.tum.de).}%
  \thanks{K. Wang was with the Chair of High-Frequency Engineering, Department of Electrical and Computer Engineering, Technical University of Munich, Germany, and is now with Infineon Technologies AG, Neubiberg, Germany.}%
\thanks{Color versions of one or more of the figures in this communication are
available online at http://ieeexplore.ieee.org.}
\thanks{Digital  Object Identifier: 10.1109/TAP.2017.2710261}
  }

\maketitle

\begin{abstract}
A novel mm-wave microstrip-fed patch antenna with broad bandwidth and wide angular coverage suitable for integration in planar arrays is designed, analyzed and verified by measurements. 
The antenna provides a bandwidth of 13.1\% between~\SI{34.1}{\giga\hertz} and~\SI{38.9}{\giga\hertz}, which is achieved by a slotted multiple resonances microstrip patch and a matching circuit in microstrip technology. 
The antenna is built on RO3003 substrate with top and ground layers, which is low cost compared to other techniques. 
For simple integration with microstrip  and frontend circuits, the feeding happens in the top layer with a microstrip coupling gap feed. 
The wide half power beamwidth is achieved by suitably designed parasitic patches for the first resonant mode. 
The second resonant mode has a wide half power beamwidth by default. 
The half power beamwidth is between~\SI{100}{\degree} and~\SI{125}{\degree} within the matched bandwidth, which is a very good value for a microstrip patch antenna radiating over a ground plane.
The measured input impedance and radiation characteristic show very good agreement with simulation results. 
\end{abstract}

\begin{IEEEkeywords}
Antenna arrays, antenna feeds, antenna pattern synthesis, microstrip antennas, mm wave communication, mobile antennas.%
\end{IEEEkeywords}

\section{Introduction} 
\IEEEPARstart{F}{or} the  next mobile communications standard 5G, mm-wave wireless communication systems are widely studied~\cite{Agiwal16}.  
To achieve high data rates, broadband antennas are required in particular at mm-wave frequencies. 
Due to the small size of mm-wave antennas, it is favorable to integrate the frontend circuit directly with the antenna  on the same substrate. 
The microstrip patch antenna is a good choice for this case. 
However, common patch antennas are often too directive for mobile applications, where a wide angular range is desired. 
For the receive case, self-mixing arrays can provide a large gain over a wide beamwidth, where the array elements can be placed sparsely, as opposed to conventional arrays~\cite{Wang15}. 
This application defines two goals for the presented antenna design, wide beamwidth and large bandwidth.

Antenna designs with a stacked setup  offer a large bandwidth~\cite{anguera2004,targonski1998}. 
Such a design is also possible at mm-wave frequencies~\cite{Waterhouse2003}. 
However, not only the antenna setup itself is complicated due to the multiple layers, also aperture coupleds feed are difficult to fabricate. 
Another common choice are patch antennas which support multiple resonances. 
One possibility feasible in a planar structure is to place parasitic patches next to a fed main patch with lower and higher resonant frequencies~\cite{kumar1985,Kumar1985v2}. 
However, this method has the disadvantage of a frequency dependent radiation characteristic. 
Another well-known patch antenna type with multiple resonances are patches with special shapes, such as the E-shaped patch antenna~\cite{wong2001,yang2001,ang2007wideband} or the C-shaped and U-slotted patch antennas~\cite{eshaped12}. 
These antennas have been reported with coaxial probe feeding and are, thus, not well suited for integration with the frontend circuits. 
Therefore, the E-shaped patch antenna has been adapted for microstrip line feeding at \SI{2.4}{GHz}~\cite{Mahgoub} and also at 60\,GHz~\cite{Jang2015,Wu2016}. 
These E-shaped patch antennas exhibit about the same, relatively large, gain of \SIrange{8}{9}{dBi} as standard rectangular microstrip patch antennas. 
 
For mm-wave communications, directive antenna arrays consisting of large gain single elements are often investigated~\cite{Sun13,Li14b,Li14,Semkin15,Ojaroudiparchin16,Khalily16}. 
Albeit, the large array gain narrows the beamwidth significantly. 
Therefore, it can be useful if the single antenna shows a wide beamwidth and a low directivity. 
Such antennas have been investigated, e.g. for positioning applications, with circular polarization achieving a beamwidth of up to \SI{140}{\degree}~\cite{Pan2014,Chen2014,Wen16}, and also with linear polarization for microstrip antennas~\cite{Chatto2009,ning2008,Khidre13,Ko2013,Ko2013b,Patel2015,Wang16TCM}. 
However, these microstrip antennas with a wide HPBW are quite narrowband and, thus, of little use in communication applications. 

In this paper, a novel multiple resonances patch antenna design with a coupling gap feed is presented which provides a wide angular range over  a large bandwidth and which is well suited for integration into planar arrays with metal backing. 
Preliminary results have already been presented in~\cite{wang_microstrip_2016}. 
This paper presents final measurement results and a detailed explanation of the underlying ideas. 
First, the broadband patch antenna design is presented and analyzed. 
Afterwards, the new technique to enhance the angular coverage of  general antennas and specifically of microstrip patch antennas by parasitic patches is introduced and implemented for the proposed antenna. 
Finally, the antenna is fabricated. The antenna input impedance and radiation pattern are verified by measurement. 

\section{Broandband Antenna Design} 

The presented mm-wave patch antenna design is implemented on RO3003 substrate with a relative permittivity of~$3$, substrate height of $h=\SI{0.254}{\milli\meter}$, which is about $0.03\lambda_0$ in terms of the free space wavelength at \SI{36}{\giga\hertz}, and copper thickness of~\SI{17.5}{\micro\meter}~\cite{RO3003}. 
For the simulation, the dielectric loss was chosen to be~$\tan\delta\approx0.004$ at~\SI{40}{\giga\hertz}. 
Compared to other techniques like  Low Temperature Cofired Ceramics (LTCC), the cost of this material is considerably lower.  Furthermore, the manufacturing process is quite simple, since no drilling, vias or multilayer-technology are required.

\begin{figure}[tp]
 \centering
  \includegraphics{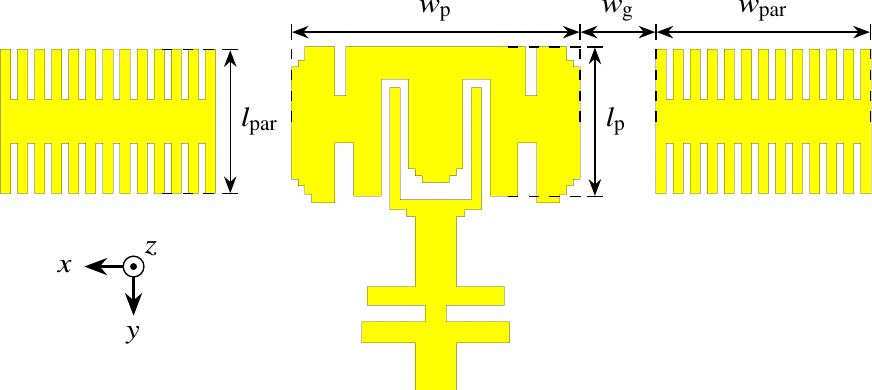}
 \caption{Complete structure of the proposed antenna including the most important dimensions.\label{fig:dims}} 
\end{figure}

\begin{figure}[tp]
 \centering
  \includegraphics[]{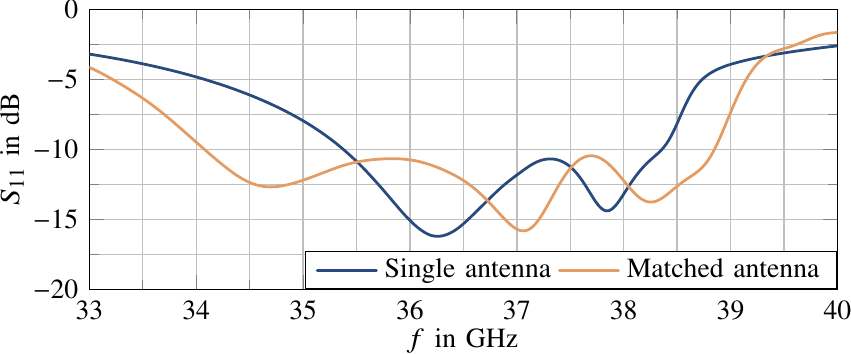}
 \caption{Comparison of the presented antenna's $S_{11}$ with and without the bandwidth-enhancing matching circuit.\label{fig:sparams_sim}} 
\end{figure}

\begin{figure}[tp]
 \centering
  \includegraphics[]{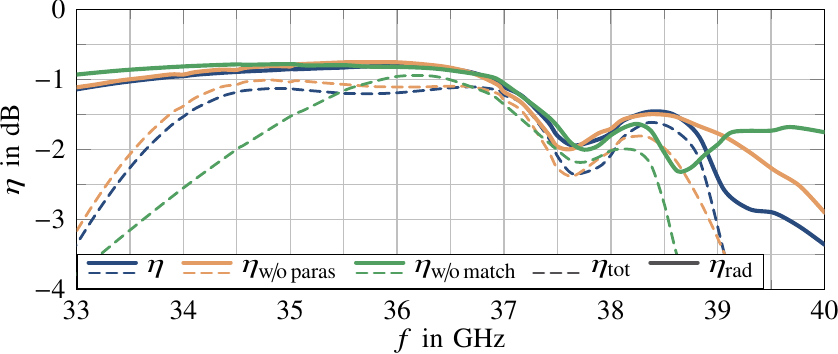}
 \caption{Comparison of the presented antenna's efficiency~$\eta$ and the same antenna design without the bandwidth-enhancing matching circuit and also without the parasitic patches.\label{fig:eff_sim}} 
\end{figure}

The broadband antenna design is achieved by several design features. 
First, as the input impedance bandwidth~\cite{Jackson1991}
\begin{equation}
B \propto \frac{hw_\mathrm p}{\varepsilon_\mathrm r \lambda_0 l_\mathrm p}
\end{equation}
is approximately proportional to the patch width $w_\mathrm p$, the general rectangular patch shape is chosen to own  a wide patch width corresponding to a large matched bandwidth. 
Furthermore, a structure with multiple resonances has been designed. 
The proposed antenna provides a~\SI{10}{\decibel} bandwidth of~8.6\%, as compared to a rectangular patch antenna on the same substrate which can provide up to~4\% relative bandwidth. 
Several shapes of patch antennas 
are known to provide a large bandwidth. 
However, they mostly require coaxial probe feeding for the excitation of their resonant modes. 
For the suggested antenna design, a structure similar to the E-shaped patch antenna, with additional slots, has been chosen such  that it can be fed in the top layer of a microstrip setup with ground plane. 
The complete antenna is shown in Fig.~\ref{fig:dims}.  
The parasitic patches depicted in Fig.~\ref{fig:dims} will be explained in the next section. 
The important dimensions of the proposed patch antenna are
\begin{alignat}{18}
&l_\mathrm p &&=&&\, \SI{2.29}{\milli\meter}\approx0.275\lambda_0,\quad&&w_\mathrm p &&=&&\, \SI{4.42}{\milli\meter}\approx0.530\lambda_0,\notag\\
&l_\mathrm{par}&&=&&\, \SI{2.21}{\milli\meter}\approx0.265\lambda_0,\quad&&w_\mathrm{par}&&=&&\, \SI{3.29}{\milli\meter}\approx0.395\lambda_0,\notag\\
&w_\mathrm g &&=&&\, \SI{1.16}{\milli\meter}\approx0.129\lambda_0.&&&&&&
\end{alignat} 
The microstrip coupling gap feed enlarges the matched bandwidth, as the antenna feed is spatially distributed~\cite[p.\,130]{Bhartia}. 
This is relevant as the current distribution of a resonant mode changes slightly over frequency as well as the current distributions of the two different resonant modes are quite different. 
Moreover, it is known that the impedance bandwidth of microstrip patch antennas can be improved by a matching network~\cite{Pues}. 
Thus, microstrip stubs are utilized to make the antenna design more broadband. 
The matching circuit changes the impedance inside the already matched frequency band only slightly, as it behaves like a flat passband filter. 
By this impedance change, the circuit is able to enhance the bandwidth by over~\SI{1}{\giga\hertz} at the lower end of the bandwidth and by~\SI{500}{\mega\hertz} at the upper end. 
Overall, an absolute~\SI{10}{\dB} bandwidth of~\SI{4.8}{\GHz} has been achieved with this antenna design. 
This relative bandwidth of 13.1\% is depicted in Fig.~\ref{fig:sparams_sim}, 
compared to~8.6\% bandwidth for the same antenna without the matching circuit. 
To verify that the power within the whole bandwidth is radiated, and not only dissipated in substrate and copper losses, the radiation efficiency $\eta_\mathrm{rad}$ and the total efficiency $\eta_\mathrm{tot}$ attained in simulation are shown in Fig.~\ref{fig:eff_sim}. %
The efficiency $\eta$ of the final antenna is compared to the efficiency~$\eta_\mathrm{w\!/\!o\,match}$ of the antenna without the enhanced matching and to the efficiency $\eta_\mathrm{w\!/\!o\,paras}$ of the antenna without the parasitic patches, whose functionality is discussed in Section III. 
The simulations for $S_{11}$ and $\eta$ and all of the following simulations have been performed in Computer Simulation Technology Microwave Studio ({CST MWS})~\cite{CST}.

\begin{figure}[tp]
 \centering
 \includegraphics{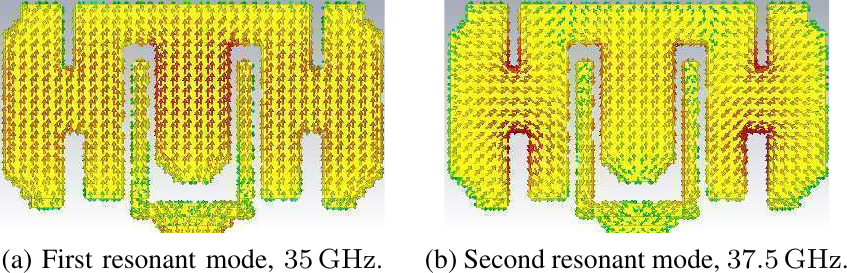}\vspace*{-0.2cm}
 \caption{Surface current distribution on the bottom of the micropstrip patch antenna for both resonant modes.\label{fig:currents}} 
\end{figure}

For a deeper insight into the working principle of the antenna, the resonant modes are analyzed. 
In Fig.~\ref{fig:currents}, the current distribution on the bottom of the top copper layer of the antenna is given for the two resonant modes. 
In Fig.~\ref{fig:currents}(a), the typical rectangular patch mode is recognized as the first resonant mode. 
In Fig.~\ref{fig:currents}(b), the first resonant mode is still excited weakly in the middle of the patch and superimposed with the current flow of the second resonant mode around the outer slots of the patch. 

\begin{figure}[!t]
 \centering
 \includegraphics{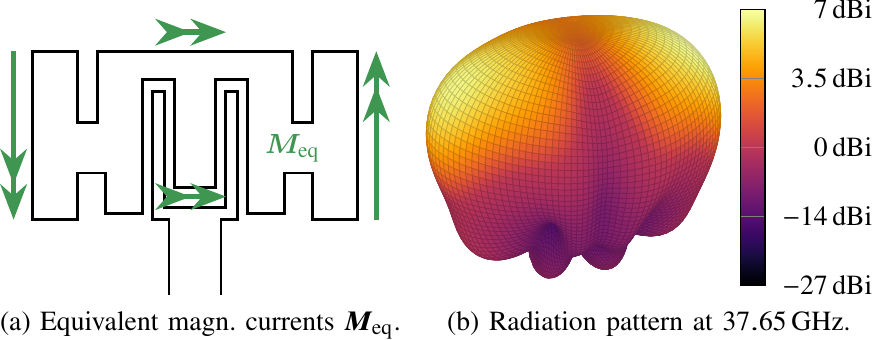}\vspace*{-0.2cm}
 \caption{Radiation analysis for the second resonant mode.\label{fig:rad_second}} 
\end{figure}

As it is well-known for the rectangular patch antenna, a radiation model can be derived utilizing equivalent magnetic currents over a ground plane~\cite{balanis2016antenna}. 
The main contribution comes from the radiating edges, which are oriented along the $x$ axis for this antenna. 
For the second resonant mode, the tangential electric field on the boundaries of the cavity also yields equivalent magnetic currents over a groundplane, however, at different locations. 
This radiation model and the corresponding {CST MWS} simulation result are given in Fig.~\ref{fig:rad_second}.  
A main contribution from the second resonant mode can be recognized with two out of phase equivalent magnetic currents, which interfere destructively in $+z$ direction. 
Hence, there are two main lobes to the side along the $x$ axis and a broad pattern is obtained. 
There is no null in $+z$ direction, as the first resonant mode is still weakly excited. 
As the structure is symmetrical and, therefore, the two beams show equal magnitude, the angular range with large power density is accordingly quite broad. 
Therefore, it makes sense to calculate the HPBW as the sum of the beamwidths in both directions. 

So far, the design goal of a broad bandwidth has been achieved by the proposed multiple resonance structure. 
It remains to enhance the beamwidth of the antenna's first resonant mode to achieve a larger angular coverage. 
This is equivalent to decreasing the directivity of the antenna. 

\section{Reducing Directivity by an Array Built With Parasitic Patches} 

\begin{figure}[!t]
 \centering
  \includegraphics[]{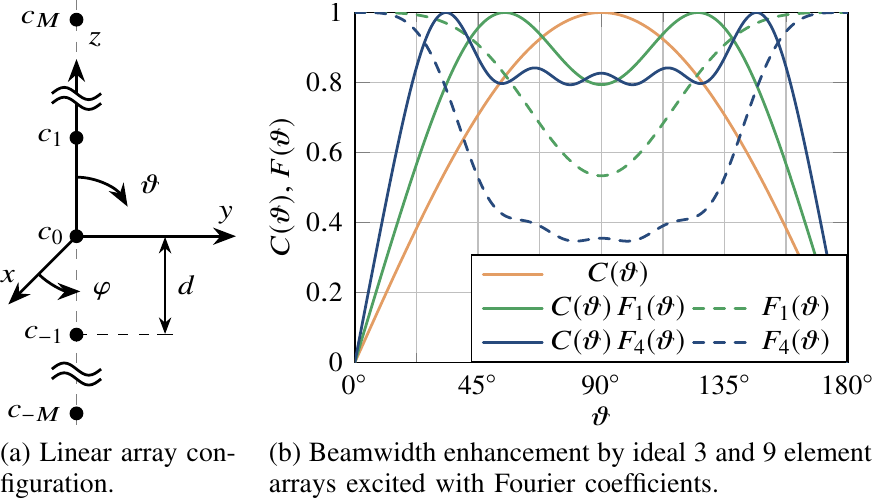}\vspace*{-0.2cm}
 \caption{Ideal array factor to reduce the directivity.\label{fig:Fourier}}
\end{figure}

The well-known resonant mode of a rectangular patch antenna, which is similar to the first resonant mode of the proposed antenna, exhibits a directive radiation pattern. 
E.g. for beamforming applications, this is not desired. 
Hence, a strategy to widen the angular range of the antenna is developed. 
The main idea is to design an array which compensates the single element radiation characteristic. 
This can be done for planar and linear array configurations by Fourier synthesis. 
A simple example is a uniformly spaced linear 1D array along the $z$ axis with nonuniform excitations. 
Such an array is depicted in Fig.~\ref{fig:Fourier}(a).  
With a total number of~$2M+1$ elements, the array factor is calculated as
\begin{equation}
F_M(\vartheta) \propto \sum\limits_{m=-M}^M c_m \mathrm{e}^{\,\mathrm{j}2\uppi m \frac{d}{\lambda}\cos\vartheta}, \label{eq:fourier-analysis}
\end{equation}
where~$d$~is the spacing in~$z$~direction and~$c_m$~is the complex excitation of the $m$th array element. 
By comparison of~\eqref{eq:fourier-analysis} with a 1D Fourier series, the array coefficients can be calculated as 
\begin{equation}
c_m  = \int\nolimits_{0}\nolimits^{\uppi}F(\vartheta) \mathrm{e}^{-\mathrm{j}2\uppi m \frac{d}{\lambda}\cos\vartheta}\sin\vartheta\mathop{}\!\mathrm{d}\vartheta, \label{eq:fourier-synthesis}
\end{equation}
to approximate a desired array factor~$F(\vartheta)$. 
This will result in a flat pattern if~$F(\vartheta)$ is chosen as the inverse of the single element radiation characteristic~$C(\vartheta)$. 
This approach assumes real valued patterns $F(\vartheta)$ and $C(\vartheta)$, as opposed to more sophisticated beam synthesis methods taking also the phase of the radiated fields into account~\cite{Echeveste2016}. 
The method is applied for a simple radiation characteristic~$C(\vartheta)=\sin\vartheta$ for $M=1$ and $M=4$ in Fig.~\ref{fig:Fourier}(b)  with an element spacing~$d={\lambda}/{2}$. 
For this example, a Fejér kernel is chosen to minimize the Gibbs phenomenon. The array factor is then given by
\begin{equation}
F_M(\vartheta) \propto \sum\limits_{m=-M}^M \left(1-\frac{|m|}{M+1}\right) \mathrm{J}_0(m\uppi) \mathrm{e}^{\,\mathrm{j}\uppi m \cos\vartheta}. \label{eq:fourier-synthesis2}
\end{equation}
This results for the three element array into
\begin{equation}
c_0=1,\qquad c_{-1}=c_1\approx-0.15.\label{eq:coeff}
\end{equation}
As it can be seen in Fig.~\ref{fig:Fourier}(b),  
the~$C(\vartheta)=\sin\vartheta$ radiation pattern shows a HPBW of~\SI{90}{\degree}. 
With the three element array, the {HPBW} is already enlargened to~\SI{120}{\degree}.
The {HPBW} is further extended to about~\SI{150}{\degree} for the nine element array. 
Thus, the benefit from the six additional antennas is small. 
In the following, only two additional elements will be utilized. 
The concept can of course be easily extended to 2D arrays, which might be even more efficient. 
To do the calculations, the array is then at best placed in the $\mathit{xy}$ plane. 

\begin{figure}[tp]
 \centering
   \includegraphics[]{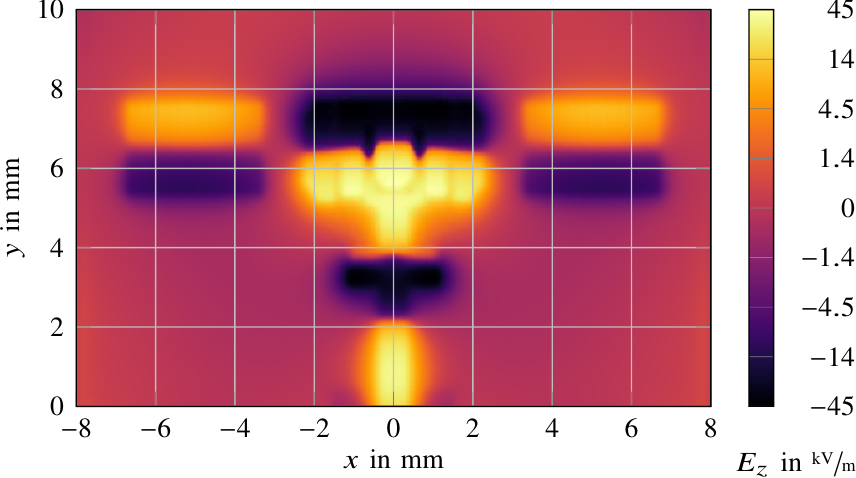}%
 \caption{Vertical electric field~$E_z$ inside the substrate at~\SI{35}{\giga\hertz}, showing the out of phase excitation of the parasitic patches.\label{fig:efield_with_parasitics}}
\end{figure}

\begin{figure}[tp]
 \centering
   \includegraphics[]{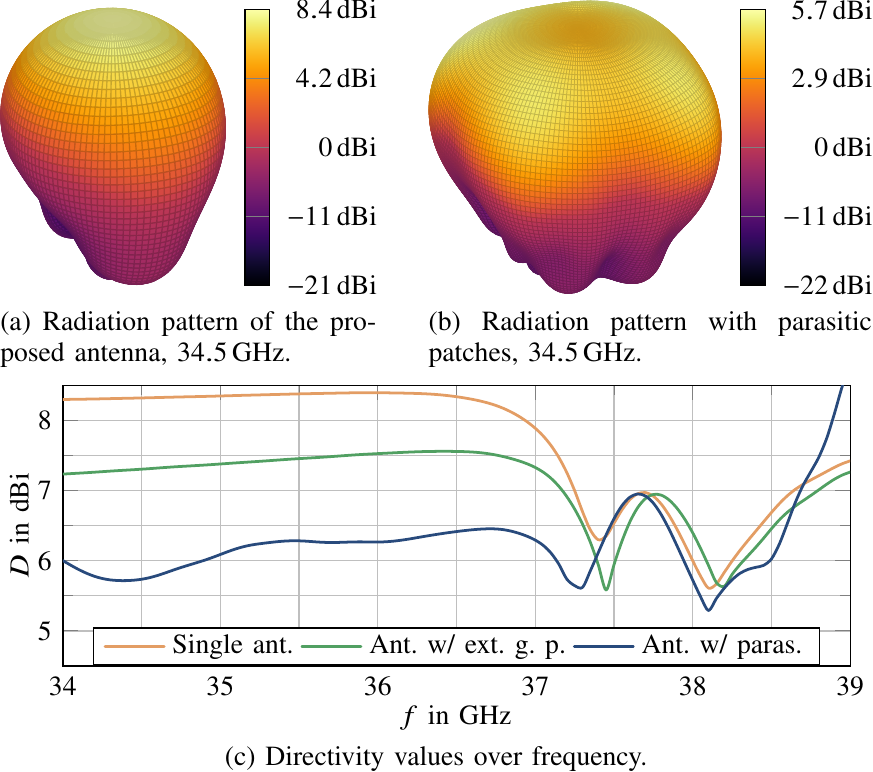}\vspace*{-0.2cm}%
 \caption{Effects of the parasitic patches and the ground plane size on the radiation behavior of the proposed antenna.\label{fig:farfield_w_wo}}
\end{figure}

The presented concept has been implemented for the rectangular patch mode of the proposed multi-resonance patch antenna. %
As generally known, the  radiation pattern of a rectangular microstrip patch antenna is broader in the $E$ plane than  in the $H$ plane. 
For the presented antenna, the $E$ plane HPBW is \SI{74}{\degree} and the $H$ plane HPBW is \SI{65}{\degree} at 36\,GHz, for a ground plane of $\SI{9}{mm}\,\times\,\SI{12}{mm}$. 
With an enlarged ground plane of $\SI{13}{mm}\,\times\,\SI{15.6}{mm}$  as employed for the parasitic patches and the matching network, these values are changed to \SI{113}{\degree} in the $E$ plane and \SI{59}{\degree} in the $H$ plane. 
Thus, it is especially suitable to broaden the radiation pattern in the $H$ plane and the additional array elements should be placed next to the non-radiating edges. 
To save the feeding network, coupling, especially by substrate waves,  is utilized to excite two parasitic patches  out of phase and with lower amplitude with respect to the main element, according to the coefficients of the three element array calculated in~\eqref{eq:coeff}. 
The distance, width and length of the parasitics are critical to achieve a broadband phase shift and excitation. 
It is found that about a quarter wavelength in the substrate is an appropriate gap width~$w_\mathrm{g}$ for the parasitic patches. 
The  width~$w_\mathrm{par}$ of the parasitics is large to provide a large bandwidth. 
The length~$l_\mathrm{par}$ is tuned such that the resonance frequency of the parasitics is lower than the one of the central patch. 
The  excitation of the parasitics is illustrated by the electric field inside the substrate in Fig.~\ref{fig:efield_with_parasitics}. 
Both lower amplitude and phase shift of the parasitic patches can be recognized. 
The influence of the parasitics on the far field pattern is depicted in Fig.~\ref{fig:farfield_w_wo}. 
For the frequency range from \SIrange{34}{37}{\giga\hertz}, the directivity is reduced from almost~\SI{8.5}{dBi} to values between~\SI{5.7}{dBi} and \SI{6.5}{dBi}. 
This corresponds to a {HPBW} enhancement in the $H$ plane from~\SI{65}{\degree} to~\SI{114}{\degree} in the {CST MWS} simulation  
at 36\,GHz. 
The $E$ plane HPBW is almost unchanged with a value of~\SI{110}{\degree}. 

The second resonant mode excited between \SI{37}{\giga\hertz} and \SI{39}{\giga\hertz} already has a broad beam, as it exhibits partial destructive interference in $+z$ direction  as seen in Fig.~\ref{fig:rad_second}. 
Therefore, it is desired that the additional parasitic patches do not influence the radiation of this resonant mode. 
It was found that slots along the $y$ direction prevent current flow in $x$ direction, as it occurs for the second resonant mode. 
Thus, parasitic patches with slots were designed which behave identical as standard rectangular patches for the first resonant mode, but show the advantegous behavior of not influencing the second resonant mode. 
The achieved directivity values over frequency are given in Fig.~\ref{fig:farfield_w_wo}(c). 
The directivity of the first resonant mode is decreased by approximately~\SI{2}{\decibel}, whereas the already low directivity of the second resonant mode remains unchanged by the choice of slotted parasitics. 
It can be concluded that the extended ground plane already reduces the directivity by widening the $E$ plane beamwidth. 
Furthermore, the parasitic patches enhance the $H$ plane HPBW in the whole bandwidth of the first resonant mode, reducing the directivity further. %
Above~\SI{38}{\giga\hertz}, a slight change in the directivity values can be recognized, as some further resonant mode of the parasitics is excited. 
Furthermore, the parasitic patches change the input impedance of the antenna at the first resonant mode and 
 the impedance bandwidth is slightly decreased by the out of phase parasitics. 
As they do not offer significant additional resonant frequencies, the known bandwidth enhancement with multiple resonators~\cite{Kumar1985v2} is clearly not observed. 

The presented design concept increases the angular coverage of a microstrip patch antenna significantly, whereas the radiation efficiency of the antenna remains unchanged. 
Thus, the proposed antenna shows all the desired properties of wide angular range, broad bandwidth and a microstrip feed. 

\begin{table}[!t]
\caption{A Comparison Between Different Antenna Performances.\label{tab:comp}}
\centering
  \includegraphics[]{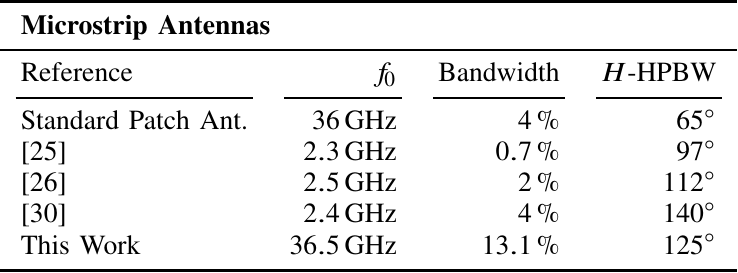}
\end{table}

The proposed antenna is compared  to other antennas in Tab.~\ref{tab:comp}, where the HPBW in the $H$ plane ($H$-HPBW), the~\SI{10}{dB} matching bandwidth and the center frequency~$f_0$ are considered. 
Only microstrip antennas built on a single-layer dielectric substrate have been chosen for comparison including the standard rectangular patch antenna. 
The given $H$-HPBWs correspond to the best possible values within the matched bandwidth, as the antenna performance varies within the matched bandwidth. 

\section{Measurement Results and Verification} 

\begin{figure}[tp]
 \centering
  \includegraphics[]{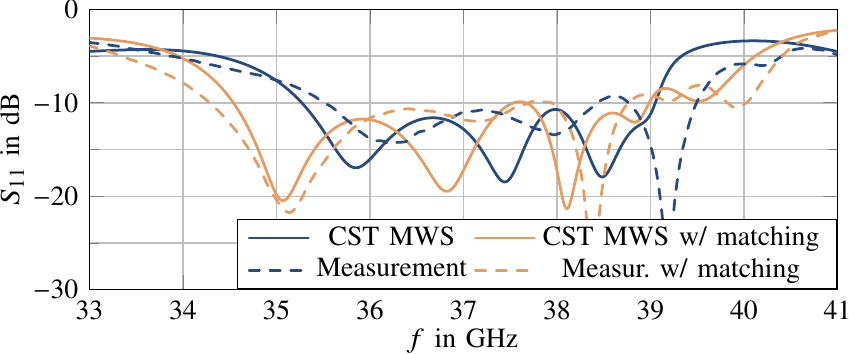}
 \caption{Comparison of simulated and measured $S_{11}$ of the proposed antenna with and without the matching circuit.}
 \label{fig:sparam_meas}
\end{figure}

To verify the proposed antenna design, several measurements have been performed. 
For the broadband design, the input impedance was measured utilizing a vector network analyzer. 
The antenna input impedance was measured with and without the presented matching circuit. 
For comparison, the coaxial connector was also added to the simulation model.
Compared to the antenna input impedance without coaxial connector in Fig.~\ref{fig:sparams_sim}, the influence of the coaxial connector is quite significant. 
The magnitude of the reflection coefficient~$S_{11}$ is given in Fig.~\ref{fig:sparam_meas} for the two simulations and the two measurements. 
The antenna without the matching stubs shows a good agreement with the {CST MWS} simulation results, except that the bandwidth changed from~\SI{3.76}{\giga\hertz} in simulation to~\SI{3.94}{\giga\hertz} in the measurement.
For the antenna with the matching circuit, the observed agreement between measurement and simulation is also pretty good. 

\begin{figure}[tp]
 \centering
 \includegraphics{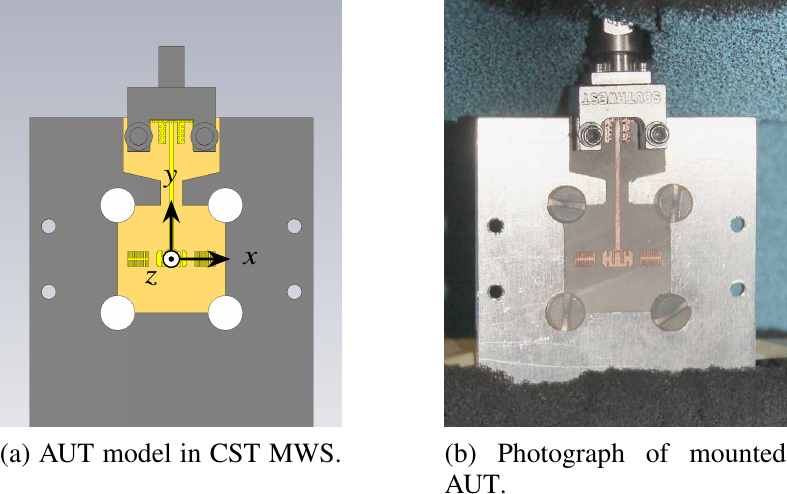}\vspace*{-0.2cm}
 \caption{Radiation pattern measurement setup.\label{fig:AUT}}
\vspace*{-0.25cm}
\end{figure}

\begin{figure}[!t]
 \centering
 \includegraphics{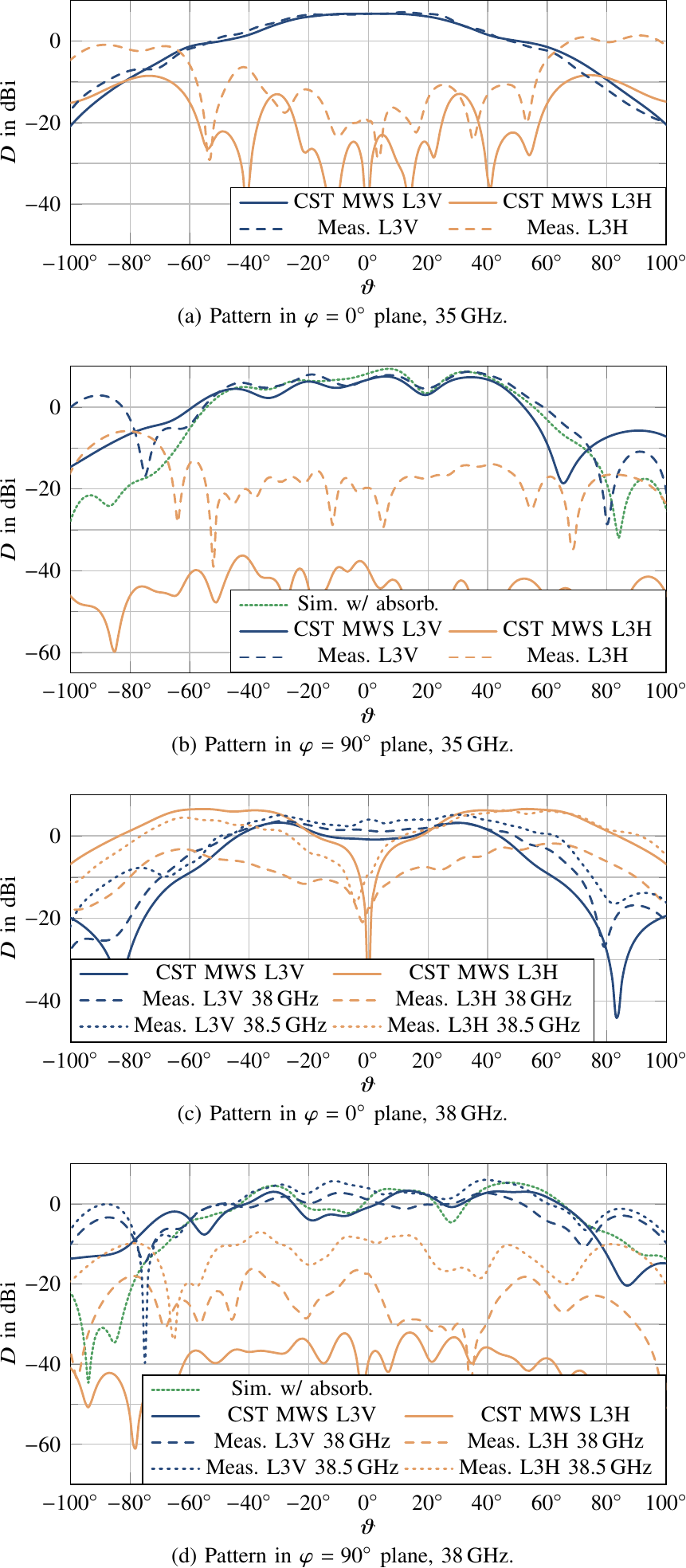}\vspace*{-0.2cm}
 \caption{Comparison of simulated and measured radiation pattern for $\varphi=\SI{0}{\degree}$ and $\varphi=\SI{90}{\degree}$ cuts at \SI{36}{\giga\hertz} and \SI{38}{\giga\hertz}, with Ludwig 3 horizontal (L3H) and vertical (L3V) polarization components.}
 \label{fig:radpatterns}
\end{figure}

The second important property of the antenna is the wide angular range of the radiation pattern. 
Hence, planar near field measurements were performed in an anechoic chamber~\cite{chamber} and the fast irregular antenna field transformation algorithm was applied to transform the measurement data into the far field, including probe correction for the measurement data~\cite{eibert2015FIAFTA}. 
To obtain $x$ and $y$ polarization components of the electric field, two measurements for co- and cross-polarization were performed, whereas the probe antenna, an open ended waveguide, was rotated by~\SI{90}{\degree} for the second measurement. 
The antenna under test ({AUT}) was placed on a metal holding plate and was fixed with polyimide screws. 
Futhermore, the pattern is influenced by the coaxial connector and nearby placed absorbers, which aim to minimize multiple interactions between {AUT} and probe. 
The {CST MWS} simulation model including the named objects and a photograph of the {AUT} are given in Fig.~\ref{fig:AUT}.   
All influences are taken into account in the simulation model, whereby the influence of the absorbers is discussed separately. 
A comparison of simulated and measured pattern cuts in the $\varphi=\SI{0}{\degree}$ and $\varphi=\SI{90}{\degree}$ plane is given in Fig.~\ref{fig:radpatterns} at~\SI{35}{\giga\hertz} for the first resonant mode and at~\SI{38}{\giga\hertz} for the second resonant mode, although the first resonant mode still has a weak contribution. 
To analyze the antenna's polarization, Ludwig 3 horizontal (L3H) and vertical (L3V) polarizations are given for the simulation and measurement results.
The limited planar scanning range leads to a valid angular measurement range of about~$\pm\SI{70}{\degree}$ in the~$\varphi=\SI{0}{\degree}$ plane and about~$\pm\SI{55}{\degree}$ in the~$\varphi=\SI{90}{\degree}$ plane. 

Both cuts at~\SI{35}{\giga\hertz} show a very good agreement between measurement and simulation in the valid angular range. 
Also for larger~$\vartheta$ values, the agreement is acceptable, although the deviations in the $\varphi=\SI{90}{\degree}$ plane are not negligible. 
The cross-polarized horizontal component shows somewhat larger values than in the simulation. 
For the second resonant mode measured at~\SI{38}{\giga\hertz}, the frequency shift of about~\SI{0.5}{\giga\hertz} observed in the scattering parameter measurement is also observed in the pattern measurement, as the excitation of the second resonant mode is shifted to higher frequencies than in the simulation. 
The second resonant mode, which gives the contribution to the cross-polarized field in the~$\varphi=\SI{90}{\degree}$ plane, is  excited less in the measurements at~\SI{38}{\giga\hertz} than in the simulation. 
The pattern agreement between~\SI{38}{\giga\hertz} in simulation and~\SI{38.5}{\giga\hertz} in measurement is good for both cuts in Fig.~\ref{fig:radpatterns}(c).  
The only exception is that the measured cross-polarized pattern is unsymmetrical, probably due to fabrication imperfections. 
Additionally, the two $\varphi=\SI{90}{\degree}$ radiation pattern might be influenced by the absorbers placed near the antenna as shown in Fig.~\ref{fig:AUT}(b).  
This has been analyzed by modeling the absorbers with $\varepsilon_\mathrm r=1.25$ and $\tan\delta=0.5$. 
The agreement between measurement and simulation in Fig.~\ref{fig:radpatterns}(b) and~\ref{fig:radpatterns}(d) is improved significantly in the range $\left|\vartheta\right|> \SI{50}{\degree}$, i.e. in the range affected by the absorbers. 
A slight disagreement around $\vartheta=\SI{-10}{\degree}$, for vertical Ludwig 3 polarization, is, however, still visible in Fig.~\ref{fig:radpatterns}(d). 

All in all, the pattern measurement and simulation results show good agreement and clearly verify the proposed multiple resonances antenna as well as the working principle of the parasitic patches. 

\section{Conclusion} 

A new   microstrip fed mm-wave patch antenna  with a bandwidth of~13.1\% was presented. 
The complexity and the costs of this antenna design are much lower than for comparable broadband patch antennas, since it is fabricated in a two layer structure with a ground plane. 
Hence, integration with frontend circuits is easily possible due the the microstrip feed. 
Furthermore, a half power beamwidth of over~\SI{100}{\degree} was achieved by emplyoing parasitics patches, which is advantegous for beamforming or general mobile applications.

\bibliographystyle{IEEEtran}
\bibliography{IEEEabrv,ref}

\end{document}